\documentclass[pre,twocolumn,nofootinbib,showpacs,superscriptaddress]{revtex4}

\usepackage{amsmath}
\usepackage{amsfonts}
\usepackage{graphicx}
\usepackage{amssymb}
\usepackage{textcomp}
\usepackage[british]{babel}
\usepackage[T1]{fontenc}
\usepackage{bm}
\usepackage{mathrsfs}
\usepackage[colorlinks=true,linkcolor=blue,citecolor=blue]{hyperref}

\begin{document}

\title{Exact first-passage time distributions for three random diffusivity models}

\author{Denis S.~Grebenkov}
\affiliation{Laboratoire de Physique de la Mati\`{e}re Condens\'{e}e (UMR 7643),
CNRS -- Ecole Polytechnique, IP Paris, 91128 Palaiseau, France}
\affiliation{Institute for Physics and Astronomy, University of Potsdam, 14476
Potsdam-Golm, Germany}
\author{Vittoria Sposini}
\affiliation{Institute for Physics and Astronomy, University of Potsdam, 14476
Potsdam-Golm, Germany}
\affiliation{Basque Center for Applied Mathematics, 48009 Bilbao, Spain}
\author{Ralf Metzler}
\affiliation{Institute for Physics and Astronomy, University of Potsdam, 14476
Potsdam-Golm, Germany}
\author{Gleb Oshanin}
\affiliation{Sorbonne Universit\'e, CNRS, Laboratoire de Physique Th\'eorique de la
Mati\`{e}re Condens\'ee (UMR 7600), 4 Place Jussieu, 75252 Paris Cedex 05, France}
\author{Flavio Seno}
\affiliation{INFN, Padova Section and Department of Physics and Astronomy "Galileo
Galilei", University of Padova, 35131 Padova, Italy}

\begin{abstract} 
We study the extremal properties of a stochastic process $x_t$ defined by a
Langevin equation $\dot{x}_t=\sqrt{2 D_0 V(B_t)}\,\xi_t$, where $\xi_t$ is
a Gaussian white noise with zero mean, $D_0$ is a constant scale factor,
and $V(B_t)$ is a stochastic "diffusivity" (noise strength), which itself
is a functional of independent Brownian motion $B_t$. We derive exact,
compact expressions for the probability density functions (PDFs) of the
first passage time (FPT) $t$ from a fixed location $x_0$ to the origin for three
different realisations of the stochastic diffusivity: a cut-off case $V(B_t)
=\Theta(B_t)$ (Model I), where $\Theta(x)$ is the Heaviside theta function;
a Geometric Brownian Motion $V(B_t)=\exp(B_t)$ (Model II); and a case with
$V(B_t)=B_t^2$ (Model III). We realise that, rather surprisingly, the FPT PDF
has exactly the L\'evy-Smirnov form (specific for standard Brownian motion)
for Model II, which concurrently exhibits a strongly anomalous diffusion. For
Models I and III either the left or right tails (or both) have a different
functional dependence on time as compared to the L\'evy-Smirnov density. In
all cases, the PDFs are broad such that already the first moment
does not exist. Similar results are obtained in three dimensions for the FPT
PDF to an absorbing spherical target.\\[0.2cm]
\end{abstract}


\maketitle

\section{Introduction}

There is strong experimental evidence that in some complex environments the
observation of a "diffusive" behaviour, i.e., of a mean-squared displacement
growing linearly with time $t$ in the form $\overline{x^2(t)}\sim t$ does
not necessarily imply that the position probability density function (PDF)
$P(x,t)$ of finding a particle at position $x$ at time $t$ is Gaussian. In
fact, significant departures from a Gaussian form have been reported, with
$P(x,t)$ having cusp-like shapes in the vicinity of $x=0$, and/or exhibiting
non-Gaussian tails. Such a behaviour was observed, e.g., for the motion of
micron-sized beads along nanotubes or in entangled polymer networks \cite{1,2},
in colloidal suspensions \cite{39} or suspensions of swimming microorganisms
\cite{44}, dynamics of tracers in arrays of nanoposts \cite{40}, transport at
fluid interfaces \cite{41,42,43}, as well as for the motion of nematodes
\cite{45}. Even more complicated non-Gaussian distributions were observed in
\textit{Dictyostelium discoideum} cell motion \cite{46,47} and protein-crowded
lipid bilayer membranes \cite{48,48a}. An apparent deviation from Gaussian
forms was evidenced in numerical simulations of particles undergoing a
polymerisation process \cite{flavio}.
 
One increasingly popular line of thought concerning the origins of such
non-Gaussian diffusion advocates a picture based on the Langevin equation
\begin{align}
\label{L}
\frac{dx_t}{dt}=\sqrt{2D_t}\,\xi_t,
\end{align}
in which $\xi_t$ is a usual white noise with zero mean and covariance
$\overline{\xi_t\xi_{t'}}=\delta(t-t')$, while the diffusivity $D_t$ is an
independent stochastic process which captures in a heuristic fashion all
possible dynamical constraints, local stimuli and interactions that a
particle may experience while moving in a heterogeneous complex environment.
 
In the pioneering work \cite{chub} Chubinsky and Slater put forth such
a random diffusivity concept for dynamics in heterogeneous systems for
which they coined the notion "diffusing diffusivity". Concretely, they
modelled the diffusivity as a Brownian particle in a gravitational field
limited by a reflecting boundary condition at $D_t=0$ in order to guarantee
positivity and stationarity of the $D(t)$ dynamics. In subsequent analyses
elucidating various aspects of the diffusing diffusivity model it was
assumed that $D_t$ is a squared Ornstein-Uhlenbeck process
\cite{jain1,jain2,Chechkin17}. Finally, \cite{tyagi} use a formulation
directly including an Ornstein-Uhlenbeck process for $D_t$. All these
models feature a stochastic diffusivity with bounded fluctuations
around a mean value, with a finite correlation time. When the process is
started with an equilibrated diffusivity distribution, the mean squared
displacement has a \emph{constant\/} effective amplitude at all times,
in contrast to non-equilibrium initial conditions \cite{Sposini2018}. The
probability density function (PDF) $P(x,t)$ of such a process is not a Gaussian
function at intermediate times.\footnote{The crossover from non-Gaussian to
Gaussian forms distinguishes the diffusing diffusivity models here from the
superstatistical approach \cite{beck} employed originally in
\cite{1,2}. In the latter case the shape of the position PDF is permanently
non-Gaussian.} Instead, $P(x,t)$ exhibits a transient
cusp-like behaviour in the vicinity of the origin and has exponential
tails.\footnote{Depending on the specific model and the spatial dimension
the exponential may have a sub-dominant power-law prefactor \cite{jain1,
jain2,Chechkin17,tyagi}.} Further extensions of this basic model were
discussed in \cite{Grebenkov2018,Sposini2018,Grebenkov2019}. Different
facets of extremal and first-passage properties were scrutinised in
\cite{Lanoiselee18,Sposini19,Grebenkov19e}. We also mention recent models
for non-Gaussian diffusion with Brownian scaling $\overline{x^2(t)}\sim
t$ based on extreme value statistics \cite{stas} and multimerisation of
the diffusing molecule \cite{eli,flavio}. Generalisations to anomalous
diffusion of the form $\overline{x^2(t)}\sim t^{\alpha}$ with $\alpha\in
(0,2)$ in terms of long-range correlated, fractional Gaussian noise was
recently discussed \cite{wei,wei1}, as well as a fractional Brownian
motion generalisation \cite{diego} of the K{\"a}rger switching-diffusivity
model \cite{kaerger}. Finally, the role of quenched disorder is analysed
in \cite{zhenya}.

We note that the diffusing diffusivity process appeared earlier in the
mathematical finance literature, where it is used for the modelling of
stock price dynamics. Indeed, if we redefine $dx_t /dt$ in \eqref{L}
as $d\ln(S_t)/dt$, we recover the celebrated Black-Scholes equation
\cite{black} for the dynamics of an asset price $S_t$ with zero-constant
trend and stochastic volatility $\sqrt{2D_t}$. In this context, the choice
of a squared Ornstein-Uhlenbeck process for $D_t$ corresponds to the Heston
model of stochastic volatility \cite{129}. The process $x_t$ in \eqref{L}
thus has a wider appeal beyond the field of transport in complex heterogeneous
media.

Several ad hoc diffusing diffusivity models in which the PDFs exhibit a
non-Gaussian behaviour for all times have been analysed recently
\cite{Sposini20} from a more general perspective, i.e., not
constraining the analysis to Brownian motion with $\overline{x^2(t)}
\sim t$ only, but also extending it to anomalous diffusion. In
reference \cite{Sposini20}, which focused mostly on power spectral
densities of individual trajectories $x_t$ of such processes---a topic
which attracted recent interest \cite{a,b,c,d}---the corresponding position
PDFs were also obtained explicitly \cite{Sposini20}.
It was demonstrated that their functional form is very sensitive to the
precise choice of $D_t$. Indeed, depending on the choice of $D_t$, one
encounters a very distinct behaviour in the long time limit: the
central part of the PDF may be Gaussian or non-Gaussian, diverge
as $|x|\to0$, or remain bounded in this limit, and also the tails may
assume Gaussian, exponential, log-normal, or even power-law forms.
 
In what follows we focus on the extremal properties of three models of
generalised diffusing diffusivity introduced in \cite{Sposini20}. In
these models particle dynamics obeys the Langevin equation \eqref{L}
with $D_t=D_0V(B_t)$, where $D_0$ is a proportionality factor---the
diffusion coefficient---while $V(B_t)$ is a (dimensionless) functional
of independent Brownian motion $B_t$, such that $B_{t=0}=0$, $\langle
B_t\rangle=0$ and
\begin{align}
\langle B_tB_{t'}\rangle=2D_B\min(t,t').
\end{align}
Here and henceforth, the angular brackets denote averaging with respect
to all possible realisations of the Brownian motion $B_t$, while the bar
corresponds to averaging over realisations of the white noise process.
We note parenthetically that the extremal properties of the Langevin
dynamics subordinated to another Brownian motion have been actively
studied in the last years within the context of the so-called run-and-tumble
dynamics. In this experimentally-relevant situation, the force acting
on the particle is a functional of the rotational Brownian motion (see
e.g., reference \cite{schehr}).

Specifically we concentrate on the first-passage times (FPTs) from
a fixed position $x_0>0$ to a target placed at the origin and determine
the full FPT PDF $H(t|x_0)$ for the following three choices for the
functional $V$:\\
(I) $V(B_t)=\Theta(B_t)$, where $\Theta(x)$ is the Heaviside theta
function such that $\Theta(x) = 1$ for $x \geq 0$, and zero,
otherwise;\\
(II) $V(B_t)=\exp(-B_t/a)$ with a scalar parameter $a$; and\\
(III) $V(B_t)=B_t^2/a^2$.\\
In Model I, which we call "cut-off Brownian motion" the process $x_t$
undergoes a standard Brownian motion with diffusion coefficient $D_0$
once $B_t>0$, and pauses for a random time at its current location
when $B_t$ remains at negative values. Albeit the mean-squared
displacement of $x_t$ grows linearly in time in this model (see
reference \cite{Sposini20}), this is indeed a rather intricate process, in
which a duration of the diffusive tours and of the pausing times have the
same \textit{broad\/} distribution. We note that this model represents an
alternative to other standard processes describing waiting times and/or
trapping events. One could think of, for instance, the comb model, in
which a particle, while performing standard Brownian motion along one
direction, gets stuck for a random time in branches perpendicular to the
direction of the relevant diffusive motion \cite{ark,ark1,ark2}.

In Model II the diffusivity $D_t$ follows so-called Geometric Brownian
Motion, as does an asset price in the Black-Scholes model \cite{black}.
Note that here the dynamics of $x_t$ is not diffusive---the process
progressively freezes when $B_t$ goes in the positive direction and
accelerates when $B_t$ performs excursions in the negative direction.
Overall the latter dominate and the mean-squared displacement exhibits
a very fast (exponential) growth with time. Lastly, in Model III the
process $x_t$ accelerates when $B_t$ goes away from the origin in
either direction, and we are thus facing again a super-diffusive
behaviour: the process $x_t$ in \eqref{L} shows a random ballistic
growth with time. In a way, such a behaviour resembles the so-called
"scaled" diffusion because for typical realisations of the process
$B_t$ one has $|B_t|\sim\sqrt{t}$ and, hence, $x_t$ evolves in the
presence of a random force whose magnitude grows with time in
proportion to $\sqrt{t}$. As shown in \cite{Sposini20} the position PDF
of this process is Gaussian around the origin and exponential in the tails.
This can be compared to scaled Brownian motion, a Markovian process with time
dependent diffusion coefficient $\mathscr{K}(t)\sim t^{\alpha-1}$ in the
ballistic limit $\alpha\to2$, whose position PDF stays Gaussian at all times
\cite{sbm}. Conversely, heterogenous diffusion processes with position
dependent diffusion coefficient $\overline{\mathscr{K}}(x)\sim|x|^{\beta}$
in the limit $\beta=1$ are also ballistic but have an exponential position
PDF (with subdominant power-law correction) at all times \cite{hdp}.

For all three models we derive exact compact expressions for the FPT PDF
$H(t|x_0)$ in one dimension and also evaluate their forms for the
three-dimensional case. We remark that, in general, the exact FPT PDFs
are known in closed-form only for a very limited number of situations (see,
e.g., \cite{Borodin,redner,schehr2,metzler,Bray13,Benichou14}, compare
\cite{denis,aljaz} for a "simple" spherical system). Thus,
our results provide novel and non-trivial examples of stochastic
processes for which the full FPT PDF can be calculated exactly.
 
The paper is outlined as follows. In section \ref{general} we present
some general arguments relating the FPT PDF and the position PDF $P(x,
t)$ for the processes governed by \eqref{L}. In section \ref{model} we
present our results for the three models under study. Finally, in
section \ref{conc} we conclude with a brief summary of our results and
an outlook.

\section{General setup}
\label{general}

A general approach for evaluating the FPT PDF for the diffusing diffusivity
models in \eqref{L} was developed in \cite{Lanoiselee18} (see also
\cite{Sposini19,Grebenkov19e}). In this approach, one takes the advantage of
the statistical independence of thermal noise and of the stochastic diffusivity
$D_t$. Qualitatively speaking, the thermal noise determines the statistics of
the stochastic trajectories of the process $x_t$, whereas the diffusing
diffusivity controls the "speed" at which the process runs along these
trajectories. As a consequence, for a particle starting from $x_0$ at time
$0$ the FPT PDF to a target, $H(t|x_0)$, can be obtained via subordination
\cite{Lanoiselee18} (see also \cite{Chechkin17}),
\begin{equation}
\label{eq:H_general}
H(t|x_0)=\int\limits_0^\infty dT \, q(t; T) \, H_0(T|x_0) ,
\end{equation}
where $H_0(T|x_0)$ is the FPT PDF to the same target for ordinary Brownian
motion with a constant diffusivity $D_0$, and $q(t;T)$ is the PDF of the
first-crossing time $\tau$ of a level $T$ by the integrated diffusivity,
\begin{equation}
\tau = \inf\{ t>0 ~:~ T_t > T\} , \qquad T_t = \int\limits_0^t dt' \, D_{t'}.
\end{equation}
In other words, $T_t/D_0$ plays the role of a "stochastic internal time" of the
process $x_t$, which relates it to ordinary diffusion. The PDF $q(t;T)$ can be
formally determined by inverting the identity \cite{Lanoiselee18}
\begin{equation}
\label{eq:q_Upsilon}
\int\limits_0^\infty dT \, e^{-\lambda T} \, q(t;T) = - \frac{\partial
\Upsilon(t;\lambda)/\partial t}{\lambda},
\end{equation}
where $\Upsilon(t;\lambda)$ is the generating function of the integrated diffusivity,
\begin{eqnarray}
\nonumber
\Upsilon(t;\lambda)&=&\left\langle \exp\left(-\lambda \int\limits_0^t dt'
\, D_{t'} \right)\right\rangle\\
&=&\left\langle \exp\left(-D_0 \lambda \int\limits_0^t
dt' \, V(B_{t'}) \right)\right\rangle.
\end{eqnarray}
In the case of a constant diffusivity, $V(z)=1$, one simply gets $\Upsilon(t;
\lambda)=\exp(-D_0\lambda t)$.

When the process $x_t$ is confined to a bounded Euclidean domain $\Omega\subset
\mathbb{R}^d$, the FPT PDF $H(t|x_0)$ can be obtained via a spectral expansion
over the eigenvalues $\lambda_n$ and eigenfunctions $u_n$ of the  Laplace operator
in which the conventional time-dependence via $e^{-D_0 t\lambda_n}$ is replaced by
$(-\partial \Upsilon(t;\lambda_n)/\partial t)$ \cite{Lanoiselee18}. In fact,
substituting the spectral expansion for ordinary diffusion \cite{redner},
\begin{equation}
H_0(T|x_0) = \sum\limits_n \lambda_n e^{-T \lambda_n} \, u_n(x_0)\int\limits_
\Omega dx \, u_n(x),
\end{equation}
into equation (\ref{eq:H_general}), one gets with the aid of (\ref{eq:q_Upsilon})
that
\begin{equation}
H(t|x_0) = \sum\limits_n \bigl(-\partial_t \Upsilon(t; \lambda_n)\bigr) \, u_n(x_0)
\int\limits_\Omega dx \, u_n(x).
\end{equation}
In turn, the analysis is more subtle for unbounded domains as one can no longer
rely on spectral expansions.

In what follows we focus on two emblematic unbounded domains, for which the FPT
probability density $H_0(T|x_0)$ for ordinary diffusion is known:\\
(i) $x_t$ evolving on a half-line $(0,\infty)$ with the starting point $x_0>0$
and a target placed at the origin, for which
\begin{equation} \label{eq:Ht0_1d}
H_0(T|x_0) = \frac{x_0 \, \exp\bigl(-x_0^2/(4T)\bigr)}{\sqrt{4\pi D_0 (T/D_0)^3}} 
\end{equation}
is the L\'evy-Smirnov distribution (with $T = D_0 t$). Substituting this function
into equation (\ref{eq:H_general}), one gets \cite{Lanoiselee18}
\begin{equation} \label{eq:Ht_1d} 
H(t|x_0) = \frac{2}{\pi} \int\limits_0^\infty \frac{dk}{k} \sin(kx_0)
\bigl(-\partial_t \Upsilon(t; k^2)\bigr).
\end{equation}
Note that the position PDF reads
\begin{equation}
\label{eq:P_1d}
P(x,t|x_0)=\int\limits_0^\infty\frac{dk}{\pi}\, \cos(k(x-x_0)) \, \Upsilon(t; k^2),
\end{equation}
such that the two PDFs are related via $\partial P(0,t|x_0)/\partial t=2\partial
H(t|x_0)/\partial x_0$. This is specific to the half-line problem.

(ii) In the second case we consider the dynamics in a three-dimensional region
outside of an absorbing sphere of radius $R$. In this case one has for ordinary
diffusion
\begin{equation}  \label{eq:Ht0_3d}
H_0^{3d}(T|x_0)=\frac{R\exp\bigl(-(|x_0|-R)^2/(4T)\bigr)}{\sqrt{4\pi D_0 (T/D_0)^3}} \,,
\end{equation}
for any starting point $x_0\in\mathbb{R}^3$ outside the target, i.e., with $|x_0|>R$.
Comparing equations (\ref{eq:Ht_1d}) and (\ref{eq:Ht0_3d}) one gets for \emph{any\/}
diffusing diffusivity process $D_t$:
\begin{equation}  \label{eq:Ht_3d}
H^{3d}(t|x_0) = \frac{R}{|x_0|}\,  H(t||x_0|-R),
\end{equation}
with $H(t|x_0)$ given by equation (\ref{eq:Ht_1d}). Note that the position PDF in
this case
\begin{equation} 
P(x,t|x_0)=\frac{1}{2\pi^2|x-x_0|}\int\limits_0^\infty dk\, k\, \sin(k|x-x_0|)
\, \Upsilon(t; k^2).
\end{equation}
We highlight that in an unbounded three-dimensional space some trajectories travel
to infinity and never reach the target, such that the target survival probability
reaches a non-zero value when time tends to infinity. This implies that the FPT
PDF is not normalised with respect to the set of all possible trajectories $x_t$.
In standard fashion, the PDF in equation \eqref{eq:Ht_3d} can be renormalised over
the set of such trajectories which do reach the target up to time moment $t$.

Using these general results, we now obtain closed-formed expressions for the FPT
PDF of models I, II, and III.

\section{Results}
\label{model}

\subsection{Model I}

We first consider the functional form $V(B_t)=\Theta(B_t)$, for which the
generating function of the integrated diffusivity can be straightforwardly
determined by taking advantage of the celebrated results due to Kac
\cite{kac} and Kac and Erd\"os \cite{erdos}. In our notations, we have
\begin{equation}
\label{mI}
\Upsilon(t;q^2)=\exp\left(-\frac{D_0q^2t}{2}\right)I_0\left(\frac{D_0q^2 t}{2}\right),
\end{equation}
where $I_\nu(z)$ is the modified Bessel function of the first kind. Note that
the inverse Laplace transform of this expression produces the celebrated L\'evy
arcsine law \cite{levy}. Curiously, this expression does \emph{not\/} depend on the
diffusion coefficient $D_B$ of Brownian motion $B_t$ driving the diffusing
diffusivity.

Substituting expression \eqref{mI} into \eqref{eq:Ht_1d} we get
\begin{equation}  \label{eq:Ht_modelI}
H(t|x_0) = \frac{x_0}{\sqrt{4\pi^3 D_0 t^3}}\exp\left(-\frac{x_0^2}{8D_0t}\right)
K_0\left(\frac{x_0^2}{8D_0 t} \right),
\end{equation}
where $K_\nu(z)$ is the modified Bessel function of the second kind. For
completeness we also provide the moment-generating function of the FPT $\mathcal{T}$,
\begin{equation}
\langle\overline{\exp(-\lambda\mathcal{T})} \rangle  
= \frac{2}{\pi}   \int\limits_{x_0 \sqrt{\lambda/D_0}}^{\infty} dz \, K_0(z) \,,
\end{equation}
where the integral can also be represented in terms of modified Struve functions.

Due to the presence of $K_\nu(z)$, the FPT PDF $H(t|x_0)$ is functionally different
from the conventional L\'evy-Smirnov probability density (\ref{eq:Ht0_1d})---we
denote it as $H^{(\mathrm{LS})}(t|x_0)$---and this difference manifests itself
both in the left and right tails of the FPT PDF. At short times $t\ll x_0^2/(8
D_0)$, (i.e., for the left tail of the FPT PDF), one gets
\begin{equation}  \label{eq:modelI_short}
H(t|x_0) \simeq \frac{\exp(-x_0^2/(4D_0t))}{\pi t}  \qquad  (t\to 0),
\end{equation}
meaning that the PDF acquires, due to the presence of $K_\nu(z)$, an additional
factor $1/\sqrt{t}$. As a consequence, $H^{(\mathrm{LS})}_0(t|x_0)/H(t|x_0)
\simeq x_0/\sqrt{D_0t}\to\infty$ in this limit, implying that $H(t|x_0)$ vanishes
faster than the L\'evy-Smirnov density.
  
\begin{figure}
\centering
\includegraphics[width=8.8cm]{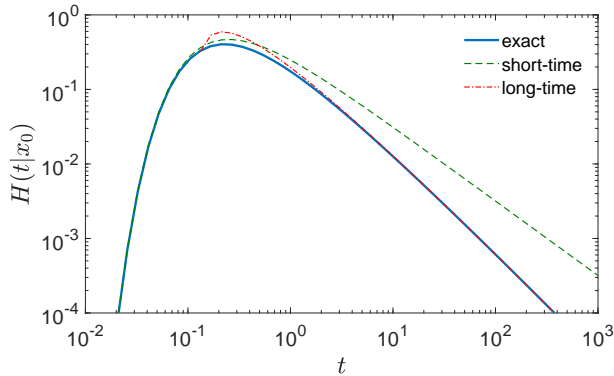}
\caption{FPT PDF $H(t|x_0)$ of the diffusing diffusivity dynamics to the
absorbing endpoint of the half-line $(0,\infty)$ for Model I. The thick
solid line shows the exact form (\ref{eq:Ht_modelI}), while the thin
dashed and dash-dotted lines present the short-time and long-time asymptotic
relations (\ref{eq:modelI_short}) and (\ref{eq:modelI_long}),
respectively. Here we set $x_0=1$ and $D_0=1$.}
\label{fig:Ht_Theta}
\end{figure}

Conversely, at long times $t\gg x_0^2/(8D_0)$ (i.e., for the right tail of the
PDF), one has
\begin{equation}  \label{eq:modelI_long}
H(t|x_0)\simeq\frac{x_0}{\sqrt{4\pi^3D_0t^3}}\left(\ln\left(\frac{16 D_0
t}{x_0^2}\right)-\gamma \right),
\end{equation}
where $\gamma=0.5772..$ is the Euler-Mascheroni constant. Hence, in the long-$t$
limit the FPT PDF of Model I due to the additional logarithmic factor has a
\textit{heavier\/} tail than the L\'evy-Smirnov density. Figure \ref{fig:Ht_Theta}
illustrates the FPT PDF and its asymptotic behaviour.   

We also note that the FPT PDF $H(t|x_0)$ resembles the free propagator of this
diffusing diffusivity motion \cite{Sposini20},
\begin{eqnarray}
\nonumber
P(x,t|x_0)&=&\frac{e^{-(x-x_0)^2/(8D_0 t)}}{\sqrt{4\pi^3 D t}}K_0\bigl((x-x_0)^2/
(8D_0t) \bigr)\\
&=&\frac{x_0}{t} \, H(t|\, |x-x_0|).
\end{eqnarray}
This curious follows equations (\ref{eq:Ht_1d}, \ref{eq:P_1d}) and from
the fact that $\Upsilon(t;\lambda)$ for this model is only a function
of $D_0t\lambda$.

\subsection{Model II}

For Model II we have $V(B_t)=\exp(-B_t/a)$ and the corresponding function
$\Upsilon(t;q^2)$ was evaluated within a different context in \cite{modII}---in
fact, $\Upsilon(t;q^2)$ is related to the moment-generating function of the
probability current in finite Sinai chains. Explicitly, $\Upsilon(t;q^2)$ is
defined by the Kontorovich-Lebedev transform
\begin{eqnarray}
\nonumber
\Upsilon(t;q^2)&=&\frac{2}{\pi} \int^{\infty}_0 dx \, \exp\left(- \frac{D_B
t}{4 a^2} x^2\right)\cosh\left(\frac{\pi x}{2}\right)\\
&&\times K_{i x}\left(2aq\sqrt{\frac{D_0}{D_B}}\right) \,,
\end{eqnarray}
where $K_{ix}(z)$ is the modified Bessel function of the second kind with
purely imaginary index. We note that the exact forms of $\Upsilon(t;\lambda)$
are also known for the case when $B_t$ experiences a constant drift
\cite{Monthus94,Oshanin12}. Inserting this expression into \eqref{eq:Ht_1d}
and performing the integrations we find that the FPT PDF of Model II is given
explicitly by
\begin{eqnarray}
\nonumber
H(t|x_0)&=&\frac{a\, {\rm arcsinh}\left(x_0/\left(2 a \sqrt{D_0/D_B}\right)\right)}{
\sqrt{\pi D_B t^3}}\\
&&\hspace*{-1.2cm}\times\exp\left(-\frac{a^2 \, {\rm arcsinh}^2\biggl(\displaystyle
\frac{x_0}{2 a\sqrt{D_0/D_B}}\biggr)}{D_B t}\right).
\end{eqnarray}
Remarkably, this is exactly the L\'evy-Smirnov density of the form
\begin{align}
\label{ls}
H(t|x_0) = \frac{X_0}{\sqrt{4\pi D_0 t^3}}\exp\left(-\frac{X_0^2}{4D_0 t}\right),
\end{align}
with an \textit{effective\/} starting point 
\begin{align}
X_0 = 2 \, a \, {\rm arcsinh}\left(x_0/\left(2 a \sqrt{D_0/D_B}\right)\right) ,
\end{align}
dependent not only on $x_0$ but also on the diffusion coefficients $D_0$
and $D_B$ in a non-trivial way.

Expectedly, the moment-generating function for the FPT is simply given by
a one-sided stable law of the form
\begin{equation}
\langle\overline{\exp(-\lambda\mathcal{T})}\rangle=\exp\bigl(-X_0\sqrt{
\lambda/D_0}\bigr),
\end{equation}
as for Brownian motion.

\subsection{Model III}

\begin{figure}
\centering
\includegraphics[width=8.8cm]{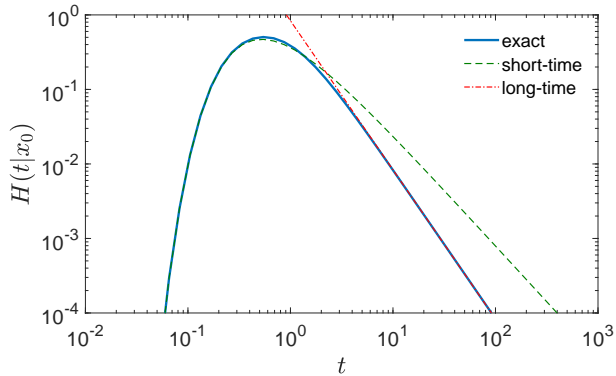}
\caption{FPT PDF $H(t|x_0)$ to the absorbing endpoint of the half-line $(0,
\infty)$ for Model III. The thick solid line shows the exact form
(\ref{eq:Ht_modelIII}) while the thin dashed and dash-dotted lines represent
the short-time and long-time asymptotics (\ref{eq:modelIII_short}) and
(\ref{eq:modelIII_long}), respectively. Here we set $x_0=1$, $D_0=1$, $D_B=1$,
and $a=1$.}
\label{fig:Ht_III}
\end{figure}

For Model III we set $V(B_t)=B_t^2/a^2$,

the function $\Upsilon(
t;q^2)$ can be calculated exactly by using the results of Cameron and Martin
\cite{88,89} (see also \cite{90})
\begin{equation}
\Upsilon(t;q^2) = \frac{1}{\sqrt{\cosh\left(c q t \right)}} \,,
\end{equation}
where $c=2\sqrt{D_B D_0/a^2}$. Inserting this expression into \eqref{eq:Ht_1d}
and performing the integral, we arrive at the rather unusual form of the FPT PDF
\begin{align}  \label{eq:Ht_modelIII}
H(t|x_0)=\frac{x_0}{\sqrt{2\pi^3}\, ct^2}\Gamma\biggl(\frac14+\frac{ix_0}{2ct}\biggr) 
\Gamma\biggl(\frac14-\frac{ix_0}{2ct}\biggr),
\end{align}
where $\Gamma(x)$ is the Gamma function. At short times, using the asymptotic
formula $|\Gamma(a+ib)|^2\simeq2\pi e^{-\pi b}/\sqrt{b}$ as $b\to\infty$ for
$a=1/4$, we get
\begin{equation}   \label{eq:modelIII_short}
H(t|x_0)\simeq\frac{2\sqrt{x_0/c}}{\sqrt{\pi t^3}}\exp(-\pi x_0/(2ct)).
\end{equation}
While the $t$-dependence of expression \eqref{eq:modelIII_short} is exactly the
same as in the L\'evy-Smirnov density, the dependence on $x_0$ is rather different,
and also the PDF depends, through the constant $c$, on the diffusion coefficient
$D_B$. In fact, setting
\begin{equation}
X_0^2 = \frac{2\pi x_0 D_0}{c} = \pi x_0 a \sqrt{D_0/D_B} \,, 
\end{equation}
we can rewrite the short-time behaviour as
\begin{equation}
H(t|x_0)\simeq\sqrt{8/\pi}\, \frac{X_0}{\sqrt{4\pi D_0t^3}}\exp(-X_0^2/(4D_0t)),
\end{equation}
which is the L\'evy-Smirnov distribution, except for the additional numerical
factor $\sqrt{8/\pi}$. At long times, the PDF exhibits the heavy tail
\begin{equation}  \label{eq:modelIII_long}
H(t|x_0) \simeq \frac{x_0 \, [\Gamma(1/4)]^2}{\sqrt{2\pi^3}\, c} \,  t^{-2} \,,
\end{equation}
i.e., it decays {\it faster\/} than the L\'evy-Smirnov distribution, but not
fast enough to insure the existence of even the first moment. Figure
\ref{fig:Ht_III} illustrates the FPT PDF $H(t|x_0)$ and its asymptotic behaviour.

\begin{figure}
\centering
\includegraphics[width=8.8cm]{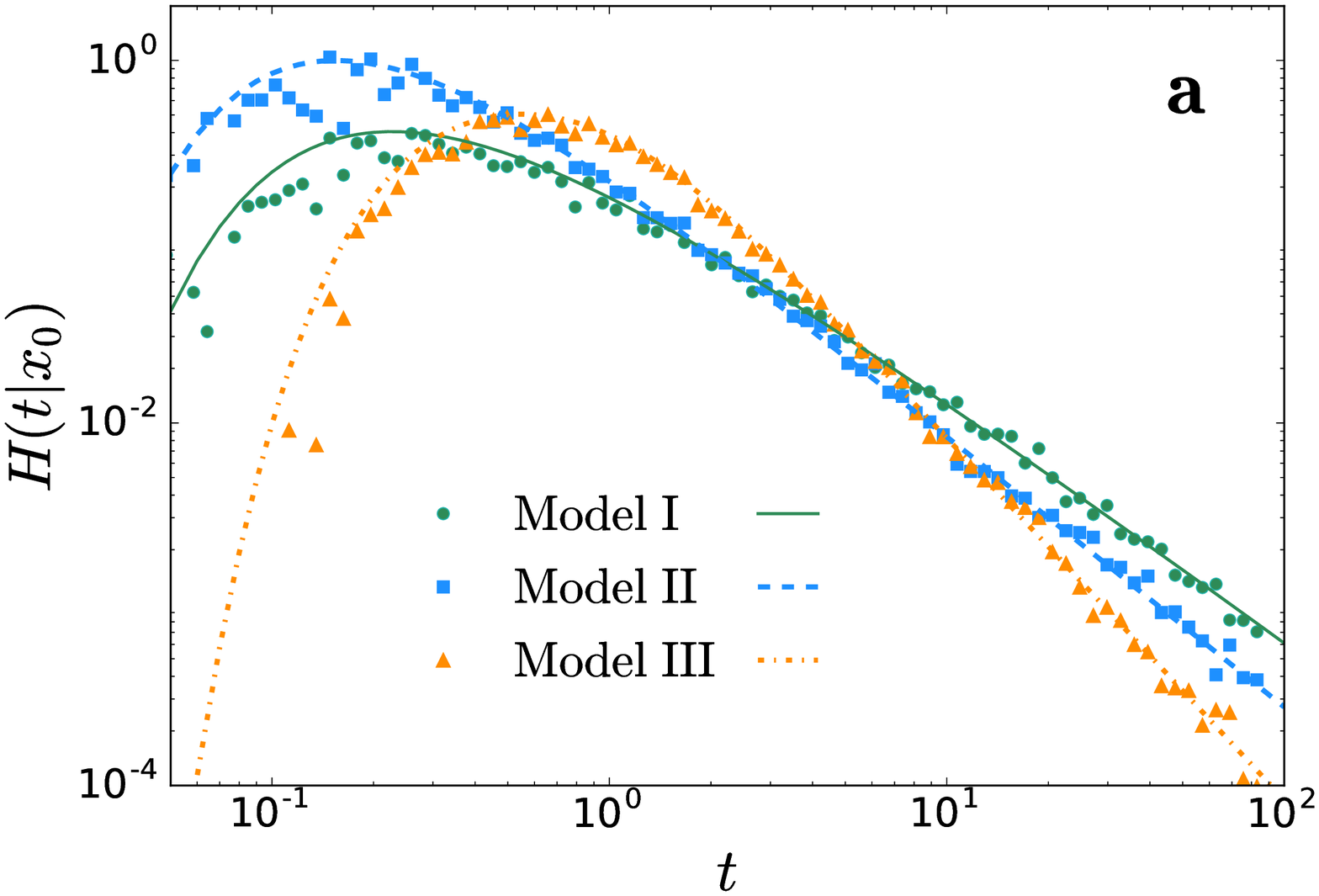}
\includegraphics[width=8.8cm]{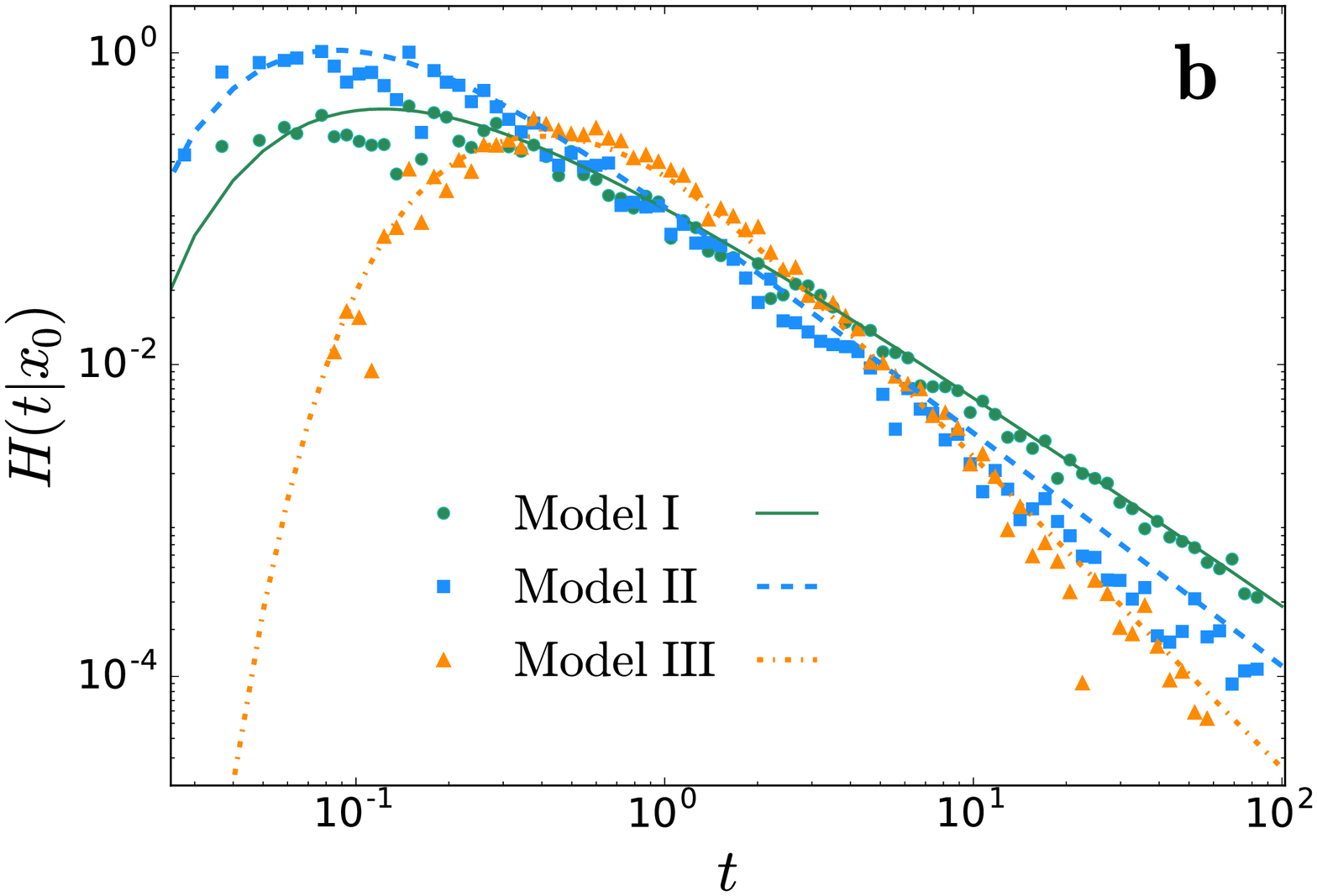}
\caption{{\bf (a)} Comparison of the FPT PDF $H(t|x_0)$ to the absorbing endpoint
of the half-line $(0,\infty)$ for the three considered models. The solid lines
represent the analytical results, while the symbols show the empirical histograms
obtained from Monte Carlo simulations with $10^4$ runs. We set $x_0=1$, $D_0=1$,
$D_B=1$, and $a=1$ (for Models II and III). {\bf (b)} Similar comparison for the
FPT PDF $H(t|x_0)$ to an absorbing sphere of radius $R=1$. The starting point is
$|x_0|=\sqrt{3}$, while the other parameters are the same.}
\label{fig:Ht_comparison}
\end{figure}

Finally, the FPT PDF $H(t|x_0)$ is plotted for the three considered models in
figure \ref{fig:Ht_comparison}(a). In addition, we present the empirical
histograms of the FPT generated by Monte Carlo simulations, observing excellent
agreement. Moreover, we depict in figure \ref{fig:Ht_comparison}(b) the FPT PDF
$H(t|x_0)$ to an absorbing sphere of radius $R$ along with Monte Carlo simulations
results.

\section{Conclusion}
\label{conc}

To conclude, we studied the extremal properties of a stochastic process $x_t$
generated by the Langevin equation \eqref{L} with a stochastic diffusivity
$V(B_t)$. The latter is taken to be a functional of an independent Brownian
motion $B_t$. For three choices of the functional form of $V(B_t)$ we derived
exact, compact expressions for the FPT PDF from a fixed initial location to the
origin. Such distributions are known only for a very limited
number of stochastic processes, and hence, our work provides novel examples
of non-trivial processes for which this type of analysis can be carried out
exactly. Similar results were obtained for the first
passage time to an absorbing spherical target in three dimensions.

Following the recent reference \cite{Sposini20}, which revealed a universal
large-frequency behaviour of spectral densities of individual trajectories
$x_t$ for the three models studied here, one could expect the {\it same\/}
short-time asymptotic behaviour of the FPT PDF for all these models, with a
generic L\'evy-Smirnov form. Indeed, in \cite{Sposini20} it was shown that
the spectral densities of individual realisations of $x_t$ decay as $1/f^2$
when $f\to\infty$, i.e., exactly as the spectral density of standard Brownian
motion. However, we realised here that the FPT PDF is of L\'evy-Smirnov
form (with an effective starting point, dependent on the diffusion
coefficients $D_B$ and $D_0$) only for Model II, in which $V(B_t)$ is
exponentially dependent on $B_t$, such that the process $x_t$ is strongly
anomalous and its mean-squared displacement grows exponentially with time.
In turn, for Model I with the cut-off Brownian motion $V(B_t)=\Theta(B_t)$,
which exhibits a diffusive behaviour $\langle\overline{x^2_t}\rangle\propto
t$, we observed essential departures from the L\'evy-Smirnov form. We saw
that the corresponding FPT PDF decays faster than the L\'evy-Smirnov law
in the limit $t\to0$, and slower than the L\'evy-Smirnov law in the limit
$t\to\infty$. For Model III with the squared Brownian motion $V(B_t)=B_t^2
/a^2$, the left tail of the FPT PDF has the L\'evy-Smirnov form with
a renormalised starting point, while the right tail decays faster. In all
models the distributions are broad such that even the first moment does not
exist.

We conclude that the universal $1/f^2$ decay of the spectral density does
not distinguish between different diffusing diffusivity models. Indeed, as
we discussed earlier, the white noise $\xi_t$ in the Langevin equation
(\ref{L}) determines the statistics of trajectories in space, whereas its
amplitude, $\sqrt{2D_t}$, can speed up or slow down the motion along each
trajectory \cite{Lanoiselee18}. The spectral density is thus more sensitive
to the spatial aspect of the dynamics, and its universal decay simply
reflects that the statistics of the trajectories governed by the white noise
is the same for all considered models. In turn, the FPT is also sensitive
to the temporal aspect of the dynamics, i.e., to the "speed", at which the
particle moves along the trajectory. This feature makes the spectral density
and the first-passage time analysis techniques complementary.

It will be interesting to extend this analysis to other models for diffusion
in heterogeneous media, in particular, when the driving noise is Gaussian
but correlated, for which the shapes of the mean squared displacement and
the position PDF were recently considered in a superstatistical approach
\cite{spako,jakub} as well as in the diffusing diffusivity picture \cite{wei,
wei1}.

\begin{acknowledgments}
D.S.G. acknowledges a partial financial support from the Alexander von
Humboldt Foundation through a Bessel Research Award. RM acknowledges the
German Science Foundation (DFG grant ME 1535/7-1) and an Alexander von
Humboldt Honorary Polish Research Scholarship from the Foundation for
Polish Science (Fundacja na rzecz Nauki Polskiej, FNP). FS acknowledges
financial support of the 191017 BIRD-PRD project of the Department of
Physics and Astronomy of Padua University.
\end{acknowledgments}

\end{document}